\documentstyle[preprint,aps,a4,12pt]{revtex}
\begin{document}
\draft
\hfill\vbox{\baselineskip14pt
            \hbox{\bf KEK-TH-607}
            \hbox{KEK Preprint 98-xxx}
            \hbox{November 1998}}
\baselineskip20pt
\vskip 0.2cm 
\begin{center}
{\Large\bf A Conjecture for possible theory for the description
of high temperature superconductivity and antiferromagnetism}
\end{center} 
\vskip 0.2cm 
\begin{center}
\large Sher~Alam\footnote{Permanent address: Department of Physics, University
of Peshawar, Peshawar, NWFP, Pakistan.}
\end{center}
\begin{center}
{\it Theory Group, KEK, Tsukuba, Ibaraki 305, Japan }
\end{center}
\vskip 0.2cm 
\begin{center} 
\large Abstract
\end{center}
\begin{center}
\begin{minipage}{14cm}
\baselineskip=18pt
\noindent
Keeping in mind some salient features of quantum symmetry groups
and in lieu of the interesting critique by Baskaran and Anderson
of the hypothesis of the SO(5) model of Zhang and the related works
of others such as Sachdev to describe superconductivity and 
antiferromagnetism we propose a conjecture to model these complex 
phenomenona by quantum groups, at least as a starting point. 
\end{minipage}
\end{center}
\vfill
\baselineskip=20pt
\normalsize
\newpage
\setcounter{page}{2}

	The Hubbard Hamiltonian [HH] and its extensions dominate
the study of strongly correlated electrons systems and the 
insulator metal transition \cite{fra91}. One of the attractive
feature of the Hubbard Model is its simplicity. It is well known
that in the HH the band electrons interact via a two-body
repulsive Coulomb interaction; there are no phonons in this model
and neither in general are attractive interactions incorporated.
With these points in mind it is not surprising that the HH
was mainly used to study magnetism. In contrast superconductivity
was understood mainly in light of the BCS theory, namely as
an instability of the vacuum [ground-state] arising from
effectively attractive interactions between electron and 
phonons. However Anderson \cite{and87} suggested that
the superconductivity in high T$_{c}$ material could arise
from purely repulsive interaction. The rationale of
this suggestion is grounded in the observation that
superconductivity in such materials arises from the
doping of an otherwise insulating state. Thus following this
suggestion the electronic properties in such a 
high T$_{c}$ superconductor material close to a 
insulator-metal transition must be considered.
In particular the one-dimensional HH is considered
to be the most simple model which can account
for the main properties of strongly correlated
electron systems including the metal-insulator 
transition. Long range antiferromagnetic order
at half-filling has been reported in the numerical
studies of this model \cite{hir89,bal90}.
Away from half-filling this model has been studied
in \cite{mor90-91}.

	Zhang \cite{zha97} proposed a unified theory
of superconductivity and antiferromagnetism 
\footnote{We note that theories of cuprates based 
upon a quantum critical point have also been suggested
by others, see for e.g. Sachdev et al. \cite{sac95}.}based
on SO(5) symmetry and suggested that there exists
an approximate global SO(5) symmetry in the low
temperature phase of the high $T_c$ cuprates.
In this model one has a five component order
parameter. Three components correspond to a spin
one, charge zero particle-hole pair condensed at the
center of mass momentum $(\pi,\pi)$, these components
represent antiferromagnetic order in the middle of 
Mott insulating state. The remaining two components
correspond  to a spin-singlet charge $\pm 2e$
Cooper pair of orbital symmetry $d_{x^2-y^2}$
condensed in zero momentum state, these last
two components are supposed to correspond to
superconductivity in the doped Mott insulator.

	Baskaran and Anderson \cite{bas97} have presented
an elegant series of criticisms of the work in ref.~\cite{zha97}. 
In our opinion these criticisms
have been stated so well that it is difficult
to improve on these, so whenever we refer to
these we stay with the original wording, almost as  
is presented in \cite{bas97}. We first state the observation 
in \cite{bas97} that is of immediate interest to us: 
\begin{itemize}
\item{}The antiferromagnetic and superconducting
phases derive from more fundamental phases, namely
from the Mott insulator and metal respectively.
Now comes the key observation, namely: ``The Mott
insulator and the metal cannot be related to a {\bf quantum
critical point}, since they differ by a local
gauge symmetry. These phases hence have no locally
stable, homogeneous intermediate phases, and cannot
be deformed continuously into each other, certainly
not by an operator as simple as an SO(5) rotation.''
\end{itemize}
In the elaboration of this point, Baskaran and Anderson
\cite{bas97} forcefully make the point that theories
of cuprates which are based on the existence of 
{\em quantum critical point} cannot realistically
describe these systems since there is no underlying
quantum critical point expressing an essential
continuity between the two phases. For further details
of this point and others raised by Baskaran and
Anderson \cite{bas97}, we refer the reader to their paper.
 	
	We now recall some useful details about quantum
groups. A quantum group can be considered as a deformation
of the classical Lie group. If we introduce a parameter
$q$, proportional $\hbar$ we may express this the idea
of this deformation as the statement  
\begin{equation}
\lim_{q \rightarrow 1}{\rm Quantum~Group}={\rm Classical~
Lie~Group}.
\label{q1}
\end{equation}
It is interesting to note some important differences
between a quantum group and an ordinary classical
group \cite{kak91}
\begin{itemize}
\item{}The braiding operations differ between the
two types of groups. In the case of a classical group
the braiding operations simply interchanges the
representations forming the tensor product of two
representations leading to +1 for a symmetric interchange
and -1 for antisymmetric interchange of the representation.
However for a quantum group the braiding operation picks
up crucial {\em phase} factors. Indeed the presence
of these nontrivial phases encourages us to consider
our proposal seriously.
\item{}One finds the relationship 
\[ q \leftrightarrow e^{2\pi i/(k+2)}\]
between the $q$ found in quantum groups and $k$ appearing 
in Kac-Moody algebra, while calculating the ``Clebsch-Gordon''
coefficients generated by these algebras \cite{kak91}.
This relationship arises due to an important reason:
unlike ordinary Lie algebras, Kac-Moody algebra contain
central charges $c$, thus even when one reduces Kac-Moody
algebra, the reduction process leads not to ordinary Lie 
algebras but to something more general, i.e. quantum groups.
\end{itemize}

	In view of the critique presented in ref.~\cite{bas97},
in particular the observation of Baskaran and Anderson 
presented above and the features of quantum groups
we are led to our conjecture: At the simplest level and as a
preliminary step, it is tempting to base a model
for superconductivity and antiferromagnetism on
a quantum group symmetry rather than the usual
classical Lie group. We consider this as
a preliminary step, since it is quite likely
that a realistic model which unifies a complex
system containing antiferromagnetic and superconducting
phases, may require the mathematical machinery
currently being used for string theory, or something even 
beyond it. Another important motivation for our
conjecture is to model Stripes. Stripes can be aptly described
as being found in the unstable two-phase region between the 
antiferromagnetic and metallic states \cite{bas97-1}. One
of the real challenges is to formulate a theory which gives rise 
to the equivalent of ``Fermi surface''\footnote{We mean, as is 
usual \cite{bas97-1}, that with a metallic state we can always
associate a ``Fermi surface'' which describes the low-energy
excitations. This is a finite volume surface in momentum space 
which arises due to all the one-particle amplitudes \cite{bas97-1}.} 
whose excitation surface derives from {\em fluctuations that 
are not uniform in space}. Intuitively one may imagine
these non-uniform fluctuations as arising out of a
nonlinear sigma-like model. We further conjecture that
superconductivity arises when two immiscible phases,
namely a 2-D antiferromagnetic state and a 3-D metallic state,
are ``forced'' to meet at $T_c \rightarrow \infty$.
As is well-known, ordinary low temperature superconductors 
arise entirely out of a ``metallic'' like  state.
In contrast High $T_{c}$ superconductors have
relatively large $T_{c}$ since we {\em cannot}
smoothly map two immiscible states [metallic and
antiferomagnetic] together. It is the lack of this
smooth mapping that is precisely responsible for
the High $T_c$. The lack of smooth mappings may
be modelled by the nontrivial phases which
arise out of the braiding operations of
quantum groups. 

\section*{Acknowledgments}
The author's work is supported by the Japan Society for
the Promotion of Science [JSPS]. 

\end{document}